\documentclass[useAMS,usenatbib]{mn2e}
\usepackage{times}
\usepackage{graphicx}
\usepackage{amsmath}
\usepackage[T1]{fontenc}
\usepackage{aecompl}

\begin{document}

\title[A No-Go Theorem for DCBHs]{A No-Go Theorem for Direct Collapse Black Holes Without a Strong Ultraviolet Background}

\author[E. Visbal et al.]{Eli Visbal$^1$\thanks{visbal@astro.columbia.edu} \thanks{Columbia Prize Postdoctoral Fellow in the Natural Sciences}, Zolt\'{a}n Haiman$^1$, Greg L. Bryan$^1$ \\ $^1$Department of Astronomy, Columbia University, 550 West 120th Street, New York, NY, 10027, U.S.A. }

\maketitle

\begin{abstract}
Explaining the existence of supermassive black holes (SMBHs) larger than $\sim 10^9 M_\odot$ at redshifts $z \ga 6$ remains an open theoretical question.  One possibility is that gas collapsing rapidly in pristine atomic cooling halos ($T_{\rm vir} \ga 10^4 \rm{K}$) produces $10^4-10^6 M_\odot$ black holes.  Previous studies have shown that the formation of such a black hole requires a strong UV background to prevent molecular hydrogen cooling and gas fragmentation.  Recently it has been proposed that a high UV background may not be required for halos that accrete material extremely rapidly or for halos where gas cooling is delayed due to a high baryon-dark matter streaming velocity.  In this work, we point out that building up a halo with $T_{\rm vir} \ga 10^4 \rm{K}$ before molecular cooling becomes efficient is not sufficient for forming a direct collapse black hole (DCBH).  Though molecular hydrogen formation may be delayed, it will eventually form at high densities leading to efficient cooling and fragmentation.  The only obvious way that molecular cooling could be avoided in the absence of strong UV radiation, is for gas to reach high enough density to cause collisional dissociation of molecular hydrogen ($\sim 10^4 ~ {\rm cm}^{-3}$) before cooling occurs.  However, we argue that the minimum core entropy, set by the entropy of the intergalactic medium (IGM) when it decouples from the CMB, prevents this from occurring for realistic halo masses.  This is confirmed by hydrodynamical cosmological simulations without radiative cooling.  We explain the maximum density versus halo mass in these simulations with simple entropy arguments.  The low densities found suggest that DCBH formation indeed requires a strong UV background. 
\end{abstract}


\section{Introduction}
Observations of high-redshift quasars imply that supermassive black holes (SMBHs) with masses larger than $\sim 10^9 M_\odot$ formed by $z=6$ \citep{2003ApJ...587L..15W,2006NewAR..50..665F, 2011Natur.474..616M}.  That such massive black holes can form within the first Gyr after the big bang presents an interesting theoretical problem \citep[for reviews see][]{2013ASSL..396..293H,2010A&ARv..18..279V}.  A seemingly natural path towards the formation of these SMBHs would be through the growth of black hole remnants from the first metal poor (Pop III) stars.  However, a $\sim 100 {\rm M_\odot}$ black hole accreting at the Eddington limit with 10 percent radiative efficiency would take roughly the age of the Universe at $z=6$ to reach $3 \times 10^9 {\rm M_\odot}$.  Radiative feedback could prevent sustained Eddington limited accretion over the required time period \citep{2007MNRAS.374.1557J,2009ApJ...701L.133A,2009ApJ...698..766M}.  Thus, stellar seeds may not have enough time to grow into the largest SMBHs observed at $z=6$.

An attractive alternative for producing the first SMBHs is direct collapse of gas in atomic cooling halos ($T_{\rm vir} \ga 10^4 \rm{K}$) into $10^4-10^6 M_\odot$ supermassive stars or quasi-stellar envelopes which quickly collapse into black holes (e.g. \citealt{2003ApJ...596...34B}; see recent reviews by \citealt{2013ASSL..396..293H,2010A&ARv..18..279V}).  This reduces the tension between the required accretion time and the age of the universe by giving black holes a head start in their mass.  The main challenge in direct collapse models is to avoid fragmentation and star formation which can occur through molecular hydrogen or metal cooling.  We note that even if fragmentation does occur it may still be possible to form a SMBH from collisions in a dense stellar cluster \citep{2008ApJ...686..801O,2012ApJ...755...81M}, however we do not address that possibility in this paper.  A strong ultraviolet (UV) background can prevent molecular hydrogen formation.  The simulations of \cite{2010MNRAS.402.1249S} show that, depending on the shape of the spectrum, a background above $J_{\rm crit} \sim 1000$ (where $J_{\rm crit}$ is in units of $10^{-21} {\rm ergs ~ s^{-1} cm^{-2} Hz^{-1} sr^{-1}}$) is required.  This critical intensity is much higher than the predicted cosmological mean \citep[see e.g.][]{2013MNRAS.432.2909F}.  Thus, DCBHs require a bright galaxy (or galaxies) a very short distance away ($\sim 10 ~ {\rm kpc}$).  Although this greatly reduces the number of dark matter halos that could host DCBHs, analytic and semi-analytic calculations still suggest that there may be enough DCBH halos to explain the abundance of SMBHs at $z=6$ \citep{2008MNRAS.391.1961D,2012MNRAS.425.2854A}.  However, black hole seeds in these models may still need to accrete at nearly the Eddington limit for a significant fraction of the age of the Universe.   

Recently DCBH models have been proposed that eliminate the need for a strong UV background.  \cite{2012MNRAS.422.2539I}  propose that shocked cold flows in atomic cooling halos can reach temperatures and densities high enough to excite the rovibrational levels in molecular hydrogen, enhancing collisional dissociation (the so-called `zone of no return').  The required density and temperature, assuming an initial ionization of $x_{\rm e}=10^{-2}$, are given by 
\begin{align}
T \ga & ~ 6000  ~ \rm{K} (n_{\rm H}/10^4 {\rm cm^{-3}})^{-1} ~ \rm{for} ~ n_{\rm H} \la 10^4 {\rm cm^{-3}}, \nonumber \\
T \ga & ~ 5000 - 6000 ~  \rm{K} ~  \rm{for} ~ n_{\rm H} \ga 10^4 {\rm cm^{-3}}.  
\end{align}
If these temperatures and densities are achieved, the halo contracts and because atomic hydrogen cooling dominates, the gas temperature stays at $T \sim 10^4 ~ \rm{K}$ preventing fragmentation.  While this is an interesting idea, recent numerical simulations \citep{2014arXiv1401.5803F} find that cold filaments shock near a halo's virial radius at relatively low density.  \cite{2012MNRAS.426.1159S} also find that molecular cooling occurs in their simulations unless the UV background is very high.

Although \cite{2014arXiv1401.5803F} find that cold flow shocks will not reach the zone of no return, they propose that it may be possible to form a DCBH if a halo grows sufficiently quickly such that it reaches the atomic cooling threshold before molecular cooling becomes efficient.  Similarly, \cite{2013ApJ...773...83X} suggest that an atomic cooling halo without stars found in their cosmological simulations may correspond to a DCBH.  Another related idea is that high baryon-dark matter streaming velocities \citep{2010PhRvD..82h3520T} could delay star formation in halos until they have grown beyond the atomic cooling threshold leading to DCBH formation \citep{2013arXiv1310.0859T}.

 In this paper, we point out that without a strong UV background, simply reaching $T_{\rm vir} \ga 10^4$ before efficient cooling occurs is not sufficient to avoid subsequent molecular cooling and fragmentation leading to the formation of a DCBH.  However, a DCBH could form in the absence of a  UV background if gas achieves a density and temperature high enough to enter the zone of no return.  To test this possibility, we run cosmological hydrodynamical simulations without radiative cooling.  Both one-zone models \citep{2001ApJ...546..635O, 2002ApJ...569..558O, 2012MNRAS.422.2539I, 2011MNRAS.418..838W} and numerical simulations \citep{2010MNRAS.402.1249S, 2014arXiv1401.5803F} have demonstrated that once efficient atomic cooling is activated outside
of the zone of no return and without a strong UV background,
molecular cooling will inevitably occur because the $\rm{H}_2$ formation timescale is shorter 
than the dynamical time.   This cooling, in turn, should lead to fragmentation.  For this reason we seek to determine if gas can reach the zone of no return before any radiative cooling (H or $\rm{H}_2$) becomes efficient.  We find that the maximum densities are several orders of magnitude smaller than the threshold required to suppress molecular cooling.  We also find that the maximum density (without radiative cooling) as a function of halo mass can be understood in terms of the core entropy.  In fact, from entropy considerations alone, we show that the zone of no return cannot be reached before efficient cooling begins.  These results support the idea that a strong UV background, or some other mechanism that continues suppressing
molecular hydrogen cooling down to high density \citep[such as enhanced heating e.g.][]{2010ApJ...721..615S} is needed for DCBH formation.

 Throughout we assume a $\Lambda$CDM cosmology consistent with the latest constraints from Planck \citep{2013arXiv1303.5076P}: $\Omega_\Lambda=0.68$, $\Omega_{\rm m}=0.32$, $\Omega_{\rm b}=0.049$, $h=0.67$, $\sigma_8=0.83$, and $n_{\rm s} = 0.96$.

\section{Simulations}
To determine the maximum gas density in $T_{\rm vir} \sim 10^4 ~ \rm{K}$ halos without atomic or molecular cooling, we ran cosmological simulations with the adaptive mesh refinement (AMR) code Enzo \citep{OShea2004, Enzo2013}.  We simulated a 1.0 comoving Mpc box starting at $z=200$ until $z=10$.  We determined the initial intergalactic medium (IGM) temperature for our adopted cosmology with Recfast \citep{1999ApJ...523L...1S} ($T_{\rm IGM}(z=200) =  480 \rm{K}$).  We performed high-resolution and low-resolution runs with $256^3$ and $128^3$ cells and particles, respectively.  Both simulations have a maximum of 8 levels of refinement corresponding to $0.015~\rm{kpc}$ (comoving) for the high-resolution run.  Our simulations require fewer levels of refinement than those with radiative cooling because the latter reach much higher densities.  To check the convergence of our high-resolution run we ran an additional $512^3$ cell/particle simulation down to $z=15$.  We used the \emph{hop} algorithm \citep{hop} implemented in yt \citep{2011ApJS..192....9T} to locate dark matter halos.

\section{Maximum Gas Density}
\begin{figure} 
\includegraphics[width=80mm]{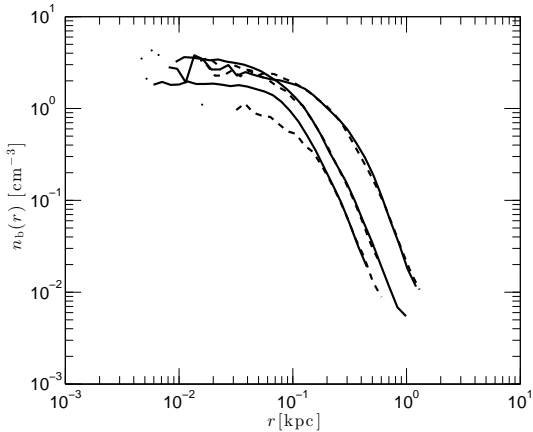}
\caption{\label{den_profiles}
Spherically-averaged gas-density profiles for a sample of dark matter halos at $z=10$.  The high-resolution and low-resolution runs are shown with solid and dashed lines respectively.  The corresponding virial masses for the profiles from right to left (for the high-$r$ portion) are $7.6 \times 10^7$, $1.4 \times 10^7$, and $5 \times 10^6  M_\odot$ .  The profiles are characterized by $n_{\rm b}(r) \propto r^{-2}$ in the outer portions and constant-density cores in the centers.  For halos with virial masses above $\sim 10^7  M_\odot$ we find reasonable convergence between the simulations.  Smaller halos have converged in the outer regions, but not the centers.}
\end{figure}

Next, we examine the distribution of gas in dark matter halos at $z=10$.  This redshift corresponds roughly to the latest cosmic time that a DCBH could form and still grow into a $\sim 3 \times 10^9 M_{\odot}$ black hole by $z=6$.  In Fig.~\ref{den_profiles}, we plot the spherically-averaged gas-density profiles for a sample of halos.  They have outer profiles given by $n_{\rm b} \propto r^{-2}$ and central constant-density cores.  We use $n_{\rm b}$ to denote the nucleon number density.  Our high-resolution simulation contains 15 halos above $M_{\rm vir} \sim10^7 M_\odot$ and 241 halos above $M_{\rm vir} \sim10^6 M_\odot$.  For halos with virial mass greater than $ \sim 10^7 M_\odot$ (which corresponds to $T_{\rm vir}=5000 ~\rm{K}$), we find the profiles look very similar for both our high-resolution and low-resolution runs.  At lower mass, the outer portions of the profiles match, but the low-resolution run has lower density cores.  We estimate that our high-resolution run has reasonable convergence at least down to $M_{\rm vir} \sim 10^6 M_\odot$ (we define the virial radius and corresponding mass as that which contains a mean density 200 times the critical cosmological density).

\begin{figure} 
\includegraphics[width=80mm]{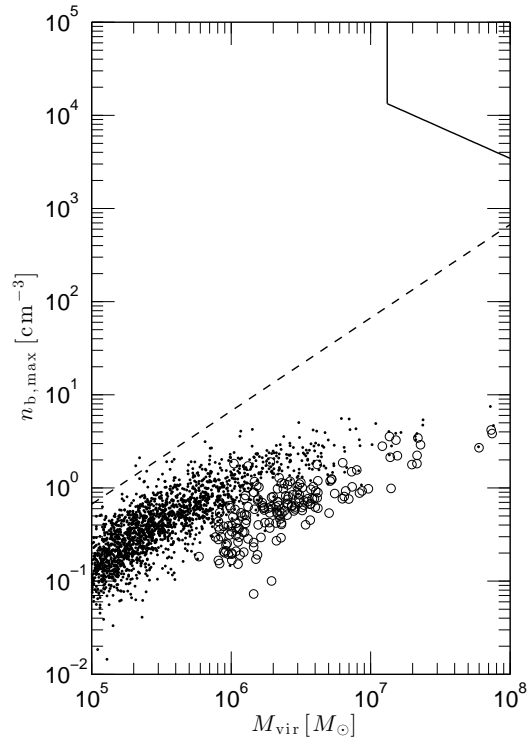}
\caption{Maximum gas density versus virial mass for halos at $z=10$.  The points are from our high-resolution run and the circles are from our low-resolution run.  The dashed line is the maximum density permitted through adiabatic contraction given by Eqn.~\ref{density_eqn}.  Low-mass halos approximately follow this scaling, while high-mass halos have nearly constant maximum density.  The zone of no return (calculated assuming gas temperature equal to $T_{\rm vir}$) is designated with solid lines in the upper right-hand corner.  It is roughly three orders of magnitude above the densities we find in our largest halos.  The left edge of this zone corresponds to $T_{\rm vir} =6000 \rm{K}$.} \label{rho_scatter}
\end{figure}

In Fig.~\ref{rho_scatter}, we plot the maximum gas density versus virial mass for the halos in our simulations at $z=10$.  We also plot the maximum density possible through adiabatic compression
\begin{equation}
\label{density_eqn}
n_{\rm b,max}(z) \sim \frac{\bar{\rho}_{\rm b}}{m_{\rm p}}\left ( \frac{T_{\rm vir}}{T_{\rm IGM}}\right )^{\frac{1}{\gamma-1}} \sim 6  \left ( \frac{T_{\rm vir}}{\rm 1000 K} \right )^{3/2} {\rm cm^{-3}}.
\end{equation}
Here $\bar{\rho}_{\rm b}$ is the universal mean baryon density, $m_{\rm p}$ is the proton mass, $T_{\rm IGM} \sim 0.012(1+z)^2$K is the temperature of the adiabatically cooling IGM, and $\gamma=5/3$ is the adiabatic index for monoatomic gas.  This equation corresponds to the entropy of the IGM when it decouples from the CMB (discussed below) and assumes the gas is heated to $T_{\rm vir}$ during virialization.  At low mass, halos roughly follow this equation, while above $\sim 3 \times 10^6 M_\odot$, the maximum density is approximately constant.  We can understand this behavior by examining the entropy profiles as discussed below.  

The maximum gas density in our halos falls roughly three orders of magnitude below that required to enter the zone of no return.  As discussed above, this indicates that DCBH formation is unlikely to proceed in the absence of a strong UV background.  Even if a halo's virial temperature is larger than $T_{\rm vir} = 10^4 \rm{K}$ before radiative cooling occurs, the density will not greatly exceed $\sim 10 ~\rm{cm}^{-3}$ at $z=10$.

\begin{figure} 
\includegraphics[width=80mm]{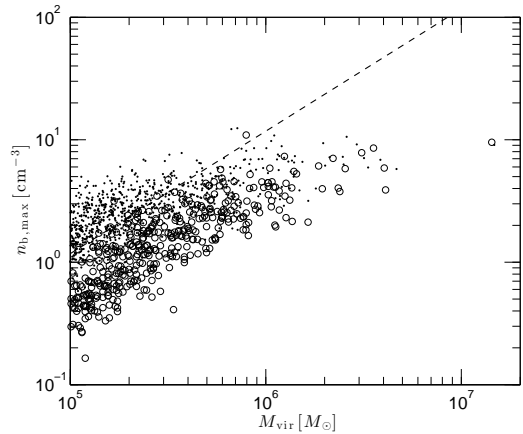}
\caption{Same as Figure~\ref{rho_scatter}, but for our high-resolution run ($256^3$ particles/cells, circles) and very high-resolution run ($512^3$ particles/cells, points) at $z=15$.  Our high-resolution run has reasonable convergence down to $M_{\rm vir} \sim 10^6 M_\odot$.  The scatter above the adiabatic limit seen at low masses is possible because the temperature in the most dense cell does not exactly match our estimated $T_{\rm vir}$. } \label{rho_scatter2}
\end{figure}

To check the convergence of these results we plot the maximum density versus virial mass at $z=15$  in Figure~\ref{rho_scatter2} and compare our high-resolution run to our very high-resolution run ($512^3$ cells/particles).  We find reasonable convergence down to roughly $M_{\rm vir} \sim 10^6 M_\odot$.

\section{Entropy Profiles}
\begin{figure} 
\includegraphics[width=80mm]{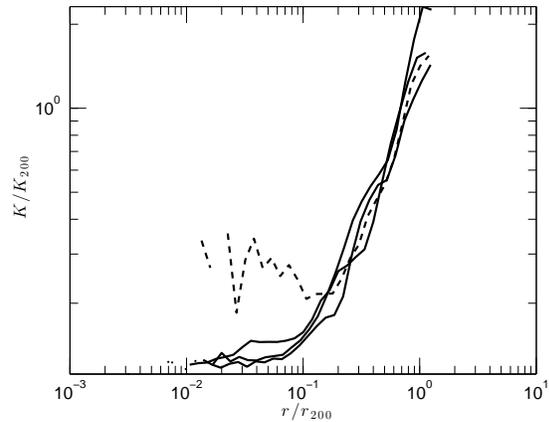}
\caption{Density-weighted spherically-averaged entropy profiles for a sample of halos from our high-resolution run at $z=10$.  For virial masses above $\sim 3 \times 10^6$ entropy profiles are nearly self similar.  This can be seen for the solid curves which have masses of $6 \times 10^7$, $1.4 \times 10^7$, and $4 \times 10^7 M_\odot$.  Smaller halos have higher central vales of $K/K_{200}$ due to the entropy floor equal to the initial entropy of the IGM.  This can been seen from the dashed curve, which is for a halo with a virial mass of $3 \times 10^6 M_\odot$. } \label{k_profiles}
\end{figure}

To better understand the maximum gas density as a function of virial mass we examine the specific entropy, defined by
\begin{equation}
\label{ent_eqn}
K = k_{\rm b} T n_{\rm b}^{-2/3},
\end{equation}
where $T$ is the gas temperature and $k_{\rm b}$ is the Boltzmann constant.  This quantity is related to the thermodynamic entropy per particle $s = \ln{K^{3/2}} + {\rm ~const} $.  We plot the density-weighted spherically-averaged entropy profile, $K(r)/K_{200}$, for a sample of our halos in Fig.~\ref{k_profiles}.  Here $K_{200}(M)=k_{\rm b} T_{\rm vir} \bar{n}_{\rm b}^{-2/3}$ is the natural entropy scale, where $\bar{n}_{\rm b}$ is $200\Omega_{\rm m}^{-1}$ times the mean cosmic baryon density.  We find that above $\sim 3 \times 10^6 M_\odot$ halos have nearly self-similar entropy profiles (i.e. the same $K(r/r_{200})/K_{200}$), charactered by a power-law outer profile and a constant core.  Self-similarity is expected because the only relevant length scale for a halo is the virial radius.  Technically the scale radius ($r_s=r_{200}/c_{\rm NFW}$) provides an additional scale, but this has a nearly one-to-one correspondence with the virial radius because the NFW concentration parameter is a weak function of mass.  Shocks and turbulence do not introduce a separate length scale because the viscous scale is much smaller than the virial radius.  Due to this self-similarity, we find that our entropy profiles look almost exactly like those from galaxy clusters simulated without radiative cooling \citep[see fig.~1 in][]{2005MNRAS.364..909V}.

\begin{figure} 
\includegraphics[width=80mm]{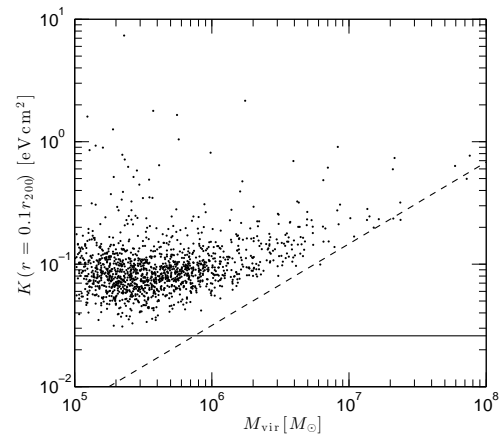}
\caption{The entropy at 10 percent of the virial radius versus virial mass for halos at $z=10$ in our high-resolution run.  The dashed line, $K \propto M^{2/3}$, is the trend expected for self-similar profiles.  The solid line is equal to the entropy of the IGM at the start of the simulation ($z=200$). The central entropy of small halos is generally slightly higher than this entropy floor.  As described in the text, this explains why the maximum gas density scales as $n_{\rm b} \propto M_{\rm vir}$ at low mass and $n_{\rm b} \propto {\rm const.}$ at high mass.  Note that the entropy at 10 percent of the virial radius is slightly higher than in the core, which contributes to the low-mass end of the distribution being slightly above the solid line.}\label{k_scatter}
\end{figure}

For halos smaller than $\sim 3 \times 10^6 M_\odot$, we find that the profiles are not self-similar, instead they have a roughy constant core entropy (i.e. $K \sim {\rm const.}$ instead of $K/K_{200}\sim {\rm const.}$ as found in larger halos).  We can understand this result in terms of a constant entropy floor.  Small dark matter halos cannot have entropy below the entropy of the IGM when it decouples from the CMB, $K_0=k_{\rm b} T_{\rm IGM}(z_0)\bar{n}_{\rm b}(z_0)^{-2/3}$, which breaks the self-similarity of $K(r)$.  In Fig.~\ref{k_scatter}, we plot the central entropy versus mass for the halos in our high-resolution simulation.  At high mass, the central regions of halos follow $K/K_{200}\sim {\rm const.}$ (dashed line), while at low mass they are just above the entropy of the IGM at the beginning of the simulation (solid line).  We note that numerical effects may be important at masses less than $\sim 10^6 ~ M_\odot$.  This picture of self-similarity at high mass and constant central entropy at low mass is analogous to entropy profiles of clusters of galaxies.  In cluster simulations without radiative cooling, preheating the IGM at early times raises the entropy profiles by a constant additive amount \citep{2007ApJ...666..647Y}.

From the behavior of the central entropy versus mass it is easy to understand our trends in central maximum density.  The central temperature of our halos follows $T \sim T_{\rm vir} \propto M_{\rm vir}^{2/3}$.  Combined with Eqn.~\ref{ent_eqn}, $K=K_0$ in small haloes gives a density equal to Eqn.~\ref{density_eqn}.  For high-mass halos, $K/K_{200} \sim 0.1$ gives a central density independent of halo mass,
\begin{equation}
\label{analytic2}
n_{\rm b,max} \approx 0.1^{-3/2} \times \bar{n}_{\rm b} = 7 \left ( \frac{1+z}{11} \right )^3 {\rm cm}^{-3}.  
\end{equation}   
This is very close to what we observe in our simulations.  For higher redshifts the maximum density in atomic cooling halos will be the smaller of Eqns. \ref{density_eqn} and \ref{analytic2}.  This results in maximum densities roughly two orders of magnitude or more below the zone of no return for $T_{\rm vir} = 10^4 ~\rm{K}$ halos at all redshifts.  We point out that our analytic maximum density estimates did not require the use of simulations (except to determine the normalization of the core $K/K_0$, which is consistent with previous galaxy cluster simulations).  Our simulations can be viewed as a confirmation of the simple entropy arguments described above.  Note that these arguments apply only to the specific case of no radiative cooling.  Radiative cooling could permit higher densities.  However, in the absence of a strong UV background this would lead to molecular cooling and fragmentation before the zone of no return is reached.

\section{Discussion}
We have shown that, without radiative cooling, gas in the cores of dark matter halos cannot reach high enough densities to enter the zone of no return.  The maximum density is limited by the minimum entropy, which is set by the entropy of the IGM when it decouples from the CMB for small halos and by the self-similarity of entropy profiles for large halos.  This is evident from our simulations.  Previous one-zone calculations and three-dimensional numerical simulations have shown that this will lead to efficient molecular hydrogen cooling and fragmentation, preventing the formation of DCBHs \citep{2001ApJ...546..635O,2010MNRAS.402.1249S}.  In fact, numerical simulations show that the zone of no return is a well defined boundary.  In the simulations of \cite{2014arXiv1401.5803F}, molecular hydrogen is artificially suppressed with a large UV background and the core density in a dark matter halo increases due to atomic hydrogen cooling.  If the artificial UV background is turned off just before the core reaches a density corresponding to the zone of no return, molecular cooling and fragmentation will still occur.  If the background is turned off after the core enters the zone of no return, the gas cannot cool below $\sim 8000 ~ \rm{K}$, potentially leading to a DCBH.

Recently it has been suggested that a DCBH could form if a halo reaches the atomic cooling threshold before molecular hydrogen cooling becomes efficient, either through rapid assembly or through delayed cooling from a high baryon-dark matter relative velocity.  In either of these scenarios, the gas density should not be significantly higher than we find in our simulations before efficient cooling.  As we explain here, cooling will proceed in the same way for fast-accreting or high-streaming velocity halos as it does for a typical atomic  
cooling halo.  The main physical difference is that the gas core could be in a deeper gravitational potential well leading to faster collapse.  However, the rate of collapse depends very weakly on dark matter halo mass.  For an NFW profile \citep{1997ApJ...490..493N}, the central dark matter density at fixed radius scales as $\rho_{\rm DM} \propto M_{\rm vir}^{1/3}$.  The one-zone baryon density is expected to follow $\frac{\rm{d}\rho_{\rm b }}{\rm{d}t} = \frac{\rho_{\rm b}}{t_{\rm dyn}}$, where $t_{\rm dyn} \propto 1/ \sqrt{\rho_{\rm b} + \rho_{\rm dm}}$.   This gives $\frac{\rm{d}\rho_{\rm b }}{\rm{d}t} \propto M_{\rm vir}^{1/6}$.  Thus, while cooling may be delayed, the physics will essentially be unaltered for halos which formed rapidly or in regions with high streaming velocity.  Since we find that gas cannot reach the zone of no return before runaway cooling begins, high streaming velocities or fast accretion in the absence of a strong UV background are not viable pathways to DCBHs.  

We note that our results do not prohibit the possibility of DCBH formation without a UV background from the trapping of Lyman-$\alpha$ radiation leading to a stiff polytropic exponent \citep{2006ApJ...652..902S, 2011MNRAS.411.1659L}, which we do not address in this paper.  We have also not considered the impact of magnetic fields, which could help to promote DCBH formation \citep[see e.g][]{2014MNRAS.440.1551L}.
  It is also possible that after gas fragmentation in the core of a dark matter halo, the resulting stars or black holes could merge through dynamical processes resulting in a SMBH.

\section{Conclusions}
Explaining the existence of $\sim {\rm a~few} \times 10^9 M_\odot$ SMBHs at $z=6$ presents an interesting theoretical challenge.  Models based on the growth of remnants from the first stars require nearly continuous Eddington limited accretion over the entire history of the universe, which seems unlikely given the expected radiative feedback.  DCBHs alleviate tension associated with this timing by forming $10^4-10^6 M_\odot$ black holes in atomic cooling halos, possibly with a short intermediate phase as a supermassive star or quasi-star.

We point out that simply delaying molecular cooling until a halo is larger than the atomic cooling threshold is not sufficient to prevent fragmentation, leading to the formation of a DCBH.  In the absence of a high UV background, molecular cooling will still occur as the gas increases in density leading to fragmentation, as determined by one-zone models and numerical simulations.  Thus, we conclude that models which produce DCBHs without a strong UV background by rapid accretion or by delayed cooling from baryon-dark matter streaming velocities are not viable.  

The only way we can envision DCBH formation without a strong UV background is if gas could reach high enough densities and temperatures to cause collisional dissociation of molecular hydrogen before the run-away process of molecular cooling can occur.  We argue that the minimum entropy of the gas will not permit halos near the atomic cooling threshold to reach these high densities.  This is confirmed by our cosmological simulations.  We find that the maximum density permitted by the entropy floor of the IGM when it decoupled from the CMB falls nearly two orders of magnitude below the zone of no return.  We note that throughout we have used the zone of no return for an initial ionization of $x_{\rm e}=10^{-2}$.  A lower initial ionization could weaken our conclusions.  However, \cite{2012MNRAS.422.2539I} tested a wide range of initial conditions and even their lowest ionization, $x_{\rm e}=10^{-5}$, still has the zone of no return above the density implied by Eqn.~\ref{density_eqn} (see their fig. 2).  Overall, our results motivate additional work on DCBH formation in the presence of a strong UV background, and the ultimate fate of a dense star cluster produced by the fragmentation process.

\section*{Acknowledgements}
EV was supported by the Columbia Prize Postdoctoral Fellowship in the Natural Sciences.  ZH was supported by NASA grant NNX11AE05G.  GLB acknowledges support from NSF grant 1008134 and NASA grant NNX12AH41G.

\bibliography{max_den_paper}
\end{document}